\documentclass[fleqn,usenatbib]{mnras}

\usepackage{newtxtext,newtxmath}

\usepackage[T1]{fontenc}

\DeclareRobustCommand{\VAN}[3]{#2}
\let\VANthebibliography\thebibliography
\def\thebibliography{\DeclareRobustCommand{\VAN}[3]{##3}\VANthebibliography}

\usepackage{graphicx}	
\usepackage{amsmath}	

\usepackage{color}

\usepackage{xcolor}
\definecolor{darkred}{RGB}{139,0,0}

\usepackage{orcidlink}

\usepackage{tabularx}
\usepackage{array}
\usepackage{xcolor}
\usepackage{color}
\usepackage{soul}
\usepackage{url}
\usepackage{graphicx}
\usepackage{amsmath}

\title[Multiwavelength analysis of EP240305a]{Multiwavelength Analysis of the Einstein Probe X-ray Transient EP240305a}

\author[Ruican Ma et al.]{
Ruican Ma $^{1} $,
Ye Li $^{2}$ \thanks{yeli@pmo.ac.cn},
Lian Tao$^{3}$ \thanks{taolian@ihep.ac.cn},
Tao An$^{4}$,
Ailing Wang$^{3}$,
Arne Rau$^{5}$,
Roberto Soria$^{6,7}$,
Huaqing Cheng$^{8}$, \newauthor
Jing Wang$^{8}$,
Hua Feng$^{3}$,
Yuanqi Liu$^{4}$,
Se\'{a}n Brennan$^{5}$,
Jingran Xu$^{9}$, 
Dave Buckley$^{10,11}$, 
Philip Charles$^{12,1,13}$, \newauthor
YuPeng Chen$^{3}$, 
Francesco Coti Zelati$^{14,15}$, 
Sebastien Guillot$^{16}$, 
Long Ji$^{17,18}$,
Chengkui Li$^{3}$ ,
Jinzhong Liu$^{19}$, \newauthor
Yuan Liu$^{8}$, 
Pierre Maggi$^{20}$, 
Itumeleng Monageng$^{10,11}$, 
Yanan Wang$^{8}$, 
Yanjun Xu$^{3}$, 
Guobao Zhang$^{21}$, \newauthor
Qingchang Zhao$^{3,22}$, 
Diego Altamirano$^{1}$,
Peter Jonker$^{23}$, 
Nanda Rea$^{14,15}$,
Xiaofeng Wang$^{24}$, 
Xuefeng Wu$^{2,25}$, \newauthor  
Weimin Yuan$^{8}$, 
Shuangnan Zhang$^{3,22}$\\
\centerline{(Affiliations can be found after the references)}
}

\date{Accepted XXX. Received YYY; in original form ZZZ}

\pubyear{\the\year{}}

\begin{document}
\label{firstpage}
\pagerange{\pageref{firstpage}--\pageref{lastpage}}
\maketitle

\begin{abstract}
We report multiwavelength observations of EP240305a, an uncatalogued X-ray transient detected by the \textit{Einstein Probe} on March 5, 2024. The source exhibits distinct characteristics across the X-ray, optical, near-infrared, and radio bands. The soft X-ray observations show two significant flares lasting $\sim 100-250$\,s, accompanied by rapid flux decay in a few days, and the optical and near-infrared data reveal a faint, candidate counterpart. In contrast, the radio observations expose a long-term spectral evolution from a self-absorbed to an optically thin state within two months, implying discrete jet ejection. We compare EP240305a with known classes of X-ray transients and find that it is unlikely to be associated with long-timescale transients such as jetted tidal disruption events or X-ray binaries. Its properties also disfavor a short-timescale stellar flare origin. 
Although the absence of optical spectroscopy prevents a redshift determination, the source exhibits properties similar to those of gamma-ray-dark gamma-ray burst-like transients, which may be associated with relativistic jets viewed off-axis or with choked jets.
The discovery of EP240305a, along with other uncataloged transients detected by the \textit{Einstein Probe}, underscores the scientific potential of highly sensitive X-ray survey telescopes and rapid-response multiwavelength follow-up observations in exploring the nature of atypical astronomical transients.
\end{abstract}




\section{Introduction} 
\label{sec:intro}
The X-ray sky hosts a diverse population of transient sources that exhibit dramatic luminosity variations over timescales ranging from milliseconds to years. 
On short timescales, Gamma-ray bursts (GRBs) rank among the most energetic events in the universe, producing X-ray flares that last from seconds to hours and reach luminosities as high as $10^{50}\,{\rm erg\,s^{-1}}$. These flares originate from relativistic jets with Lorentz factors of $\Gamma \sim 100$, launched by different progenitor systems: long GRBs typically arise from the core collapse of massive stars \citep{Woosley2006}, whereas short GRBs originate from the mergers of compact object binaries \citep{Berger2014}. However, the precise nature of their progenitor systems \citep{Levan2016} and the mechanisms that drive their central engines, whether through the collapsar or magnetar model \citep{Nagataki2018}, remains uncertain. Beyond GRBs, neutron star low-mass X-ray binaries (NS-LMXBs) can also exhibit recurrent X-ray bursts lasting from seconds to minutes. These bursts fall into two categories: Type I bursts, which result from unstable thermonuclear burning of accreted hydrogen and/or helium on the neutron star’s surface, and Type II bursts, which originate from accretion instabilities within the disk \citep{Galloway2008, Galloway2021}. Another class of highly luminous transients arises from magnetar giant flares, where young neutron stars with extreme magnetic fields $10^{14}-10^{15}$\,G release energy at rates of up to $10^{47}\,{\rm erg\,s^{-1}}$ within fractions of a second, and are often followed by pulsating tails that persist for several hundred seconds \citep[e.g.,][]{Mereghetti2008, Turolla2015}.  

Long-timescale X-ray transients arise from accretion episodes.
Among them, tidal disruption events (TDEs), where a supermassive black hole disrupts and accretes a passing star, produce X-ray emission that persist for months to years. The observed luminosities of these events span from $10^{41}\,{\rm erg\,s^{-1}}$ in non-jetted cases, characterized by thermal spectra, to $10^{47}\,{\rm erg\,s^{-1}}$ when relativistic jets are present, resulting in non-thermal emission \citep[see][for a review]{Gezari2021}. In black hole X-ray binaries (XRBs), for example, outbursts lasting weeks to months are driven by thermal-viscous instabilities in the accretion disk \citep{Lasota2001}, with typical luminosities ranging from $10^{35}\,{\rm erg\,s^{-1}}$ to $10^{39}\,{\rm erg\,s^{-1}}$ \citep[see][for a review]{Tetarenko2016}. The variability of these systems, ranging from seconds to months, reflects accretion-driven evolution, shedding light on compact object dynamics as well as strong gravity effects.

On March 5, 2024, during the pre-performance verification phase, the \textit{Einstein Probe} \citep[\textit{EP;}][]{Yuan2016, Yuan2018, Yuan2022, Yuan2025} identified a new, previously uncatalogued transient source, EP240305a, characterized by two short soft X-ray flares \citep{atel16509}. Follow-up observations were conducted across X-ray, optical, and radio wavelengths.
In this work, we present the multiwavelength results. The paper is structured as follows: Section~\ref{sec:data_reduction} describes the observations and data reduction, Section~\ref{sec:results} presents the results from the X-ray, optical, and radio data, and the properties of EP240305a are discussed in Section~\ref{sec:disc_conclu}.

\section{Observations and Data Reduction} 
\label{sec:data_reduction}

\subsection{\textit{EP}} 

The \textit{EP} \citep{Yuan2016, Yuan2018, Yuan2022, Yuan2025} is an interdisciplinary X-ray observatory dedicated to time-domain astronomy and high-energy astrophysics, led by the Chinese Academy of Sciences (CAS) in a collaboration with the European Space Agency (ESA), the Max-Planck Institute for Extraterrestrial Physics (MPE) in Germany, and the Centre National d’Etudes Spatiales (CNES) in France. The \textit{EP} satellite was launched on January 9, 2024, and carries two instruments: the wide-field X-ray telescope \citep[WXT;][]{Cheng2025} and the follow-up X-ray telescope \citep[FXT;][]{Chen2020}. 
\textit{EP}/WXT, equipped with the novel lobster-eye micro-pore optics, is designed to monitor the soft X-ray (0.5--4 keV) sky with a high sensitivity of $\sim 1$ mCrab ($\sim (2-3) \times 10^{-11}~{\rm erg~s^{-1}~cm^{-2}}$ in 0.5--4 keV) within a $\sim1000~{\rm s}$, at a high cadence (the entire night sky can be generally covered every $\sim5$ hours).
This sensitivity surpasses other currently operating wide-field monitors by a factor of several tens of times.
\textit{EP}/FXT is a traditional Wolter-I type X-ray telescope, designed to perform rapid follow-up observations for the X-ray transients discovered by \textit{EP}/WXT. It provides source localization of $5-10''$ and sensitivity of $\sim10^{-14}~{\rm erg~s^{-1}~cm^{-2}}$ in the 0.3--10\,keV band for a $10^4~{\rm s}$ exposure.

EP240305a was first identified by the \textit{EP}/WXT as a new transient source at R.A. (J2000) $=$ 08$^{\rm h}$11$^{\rm m}$37$^{\rm s}$, Dec.(J2000) $=\ -54^{\circ}$39\arcmin25$\farcs$, with a positional uncertainty of 3 arcmin (90\% confidence level). This initial detection was made by CMOS sensor No. 19 during observation ObsID 13600005096, which commenced on 2024 March 5 at 12:27:27 UT (MJD 60374.5), with a total exposure time of 17.7\,ks. 
The data were processed utilizing the {\sc wxtpipeline} tool, version 0.1.0. The energy spectrum and light curve of the source were extracted using a circular region with a radius of 9\farcm2. An identical circular region, placed sufficiently far from the source to avoid contamination, was used to extract the background.

\textit{EP}/FXT conducted a follow-up observation on 2024-03-26T11:04:50 (MJD 60395.5) in Full Frame (FF) mode, with ObsID 08503145728 and an exposure time of 9.5\,ks. 
The data processing was performed using the {\sc fxtchain} tool from the FXT Data Analysis Software \citep[\texttt{FXTDAS;}][]{2025ZhaoFXTDAS}, with calibration version CALDB\_v1.05. 
Since the source was not significantly detected in the \textit{EP}/FXT observation, we used the source position reported by \textit{Swift}/XRT (see Section~\ref{data_ana:xrt}) to derive an upper limit on the source flux.
Source and background counts were extracted with the {\sc xselect} tool for the upper-limit calculation, using a circular region of 70{\arcsec} radius centered on the source position. A background region with the same radius was selected as far from the source as possible. The ARF and RMF files were generated using the \texttt{fxtarfgen} and \texttt{fxtrmfgen} commands, respectively.

\subsection{\textit{Swift}/XRT} 
\label{data_ana:xrt}
The \textit{Neil Gehrels Swift Observatory} X-ray Telescope \citep[\textit{Swift/XRT};][]{Burrows2005} performed two observations (ObsIDs 00016561001 and 00016561002) on MJD 60376.6 and 60378.6, with exposure times of 2.7\,ks and 1.0\,ks, respectively. Both observations were carried out in PC mode due to the low count rate of the source. 
The source position was determined from \textit{Swift}/XRT observations using the \textit{Swift}/XRT products generator\footnote{\url{https://www.swift.ac.uk/user_objects/}} \citep{Evans2009}, yielding R.A. (J2000) $=$ 08$^{\rm h}$11$^{\rm m}$40$^{\rm s}$.91 and Dec. (J2000) $=\ -54^{\circ}$39{\arcmin}09{\farcs}2, with a 5{\farcs}6 uncertainty (90\% confidence level). The spectra and corresponding response files were also extracted using the same \textit{Swift}/XRT products generator.
The data reduction used software version {\sc swxrtdas} v3.7.0 and CALDB version x20230725. The default PC mode grade (0--12) was applied. We ignored the data below 0.3\,keV and rebinned the spectra to have at least 1 count per energy bin.

\subsection{\textit{GROND}} 
Observations with the seven-channel Gamma-Ray burst Optical Near-infrared Detector \citep[\textit{GROND};][]{Greiner2008} mounted at the MPG 2.2\,m telescope at ESO’s La Silla observatory were performed on MJD 60377.102, 60379.045, and 60598.343. In the first two nights, 8\,min integrations were obtained simultaneously in the $J$, $H$, and $K_{\rm s}$ bands. On the third night, also the optical g$^{\prime}$, r$^{\prime}$, and i$^{\prime}$ bands were available, and the exposure in all bands was $\sim30$\,min. The data were reduced using the standard IRAF-based GROND pipeline \citep{Kruehler2008}. 

\subsection{\textit{SVOM/VT}} 
We have observed the field of EP240305a with the Visible Telescope \citep[VT,][]{Fan2020} onboard the SVOM satellite\footnote{SVOM, launched on 2024, June 22, is a Chinese-French space mission dedicated to the detection and study of gamma-ray bursts. Please see \citet{Atteia2022} and the white paper given by \citet{Wei2016} for details.} on MJD 60732. The observation lasted for 4 orbits in both VT\_B $\mathrm{(400-650nm)}$ and VT\_R $\mathrm{(650-1000nm)}$ channels simultaneously. The exposure time is 60 seconds for each frame. A total of 46 frames were obtained in each  channel during the snapshot. 

The raw images were reduced by the dedicated pipelines through standard procedures, including bias subtraction, flat-field correction and dark current removement. For each channel, the reduced images were combined after astrometric calibration.

\subsection{ATCA} 

EP240305a was observed over six epochs with the Australia Telescope Compact Array (ATCA) in the 6A configuration over 81 days (project code CX563, and C3615), beginning one week after its discovery by \textit{EP}/WXT \citep{2024ATel16555....1A}. The most precise localisation provided by ATCA is R.A. (J2000) $=$ 08$^{\rm h}$11$^{\rm m}$40$^{\rm s}.88$, Dec.(J2000) $=\ -54^{\circ}$39$^{\prime}$07$\farcs 46$, with a positional uncertainty of $0.06 \times 0.02$ arcsec. ATCA consists of six 22-meter antennas with a maximum baseline of 6 km, enabling high-resolution radio observations. For amplitude and band-pass calibration, 0823--500 was used, while this bright source is also used as the gain and phase calibrator. The Compact Array Broadband Backend (CABB) was used for L/S and C/X band observations, centered at three frequencies, 2.1 GHz, 5.5 GHz, and 9 GHz, each with a bandwidth of 2 GHz. We use the Common Astronomy Software Applications \citep[CASA, version 6.2.1.7,][]{2007ASPC..376..127M} for the data processing. The data reduction followed the CASA standard calibration pipeline, which included manual and automated radio frequency interference (RFI) flagging, bandpass and absolute flux calibration, and time-dependent gains calibration. Imaging employed multi–frequency synthesis with Briggs weighting.

\section{Analysis and Results} 
\label{sec:results}

This section details the multiwavelength results of EP240305a, including X-ray data gathered by \textit{EP}/WXT, \textit{EP}/FXT, and \textit{Swift}/XRT, optical and near-infrared data captured by \textit{GROND}, and \textit{SVOM}/VT, as well as radio data from ATCA.

\subsection{X-ray properties}
\label{sec:xray}

\subsubsection{Light curve}
\label{sec:xray-lc}

\begin{figure*}
\centering
\includegraphics[width=\textwidth]{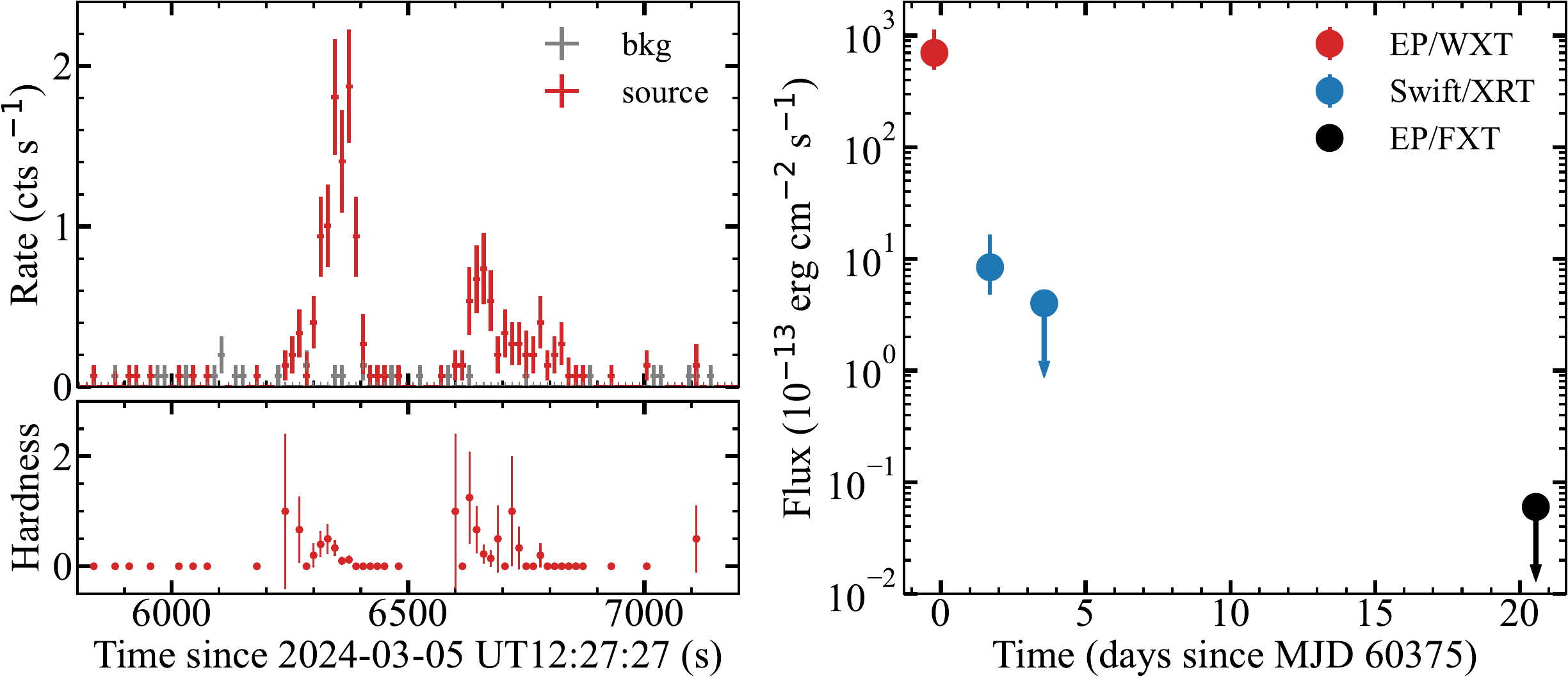} \\
\caption{Top left panel: \textit{EP}/WXT count rate for EP240305a in the 0.5--4 keV energy band with 15\,s time bins. The source is shown in red and the background is in gray. Two notable flares are evident. Bottom left panel: Hardness ratio ((2--4)\,keV/(0.5--2)\,keV) of the two flares. Right panel: Unabsorbed flux of EP240305a in the 0.5--10 keV energy band, as observed by \textit{EP}/WXT in red, \textit{Swift}/XRT in blue, and \textit{EP}/FXT in black.
\label{fig:lc}
}
\end{figure*}

As EP240305a was first detected by \textit{EP}/WXT, we extracted its light curve using 15\,s bins (Fig.~\ref{fig:lc}, top-left panel). The observation reveals two distinct flares. The first flare lasted approximately 120\,s and reached a peak count rate of 1.8\,$\rm{cts\,s^{-1}}$, corresponding to a 0.5--4\,keV flux of $3\times10^{-9}\,\rm{erg\,cm^{-2}\,s^{-1}}$ (for details of the spectral used model to calculate the flux, see Section~\ref{sec:xray-pha}). About 200\,s later, a second, weaker flare occured, lasting around 250\,s with a shallower decaying phase. It peaks at a count rate of 0.6\,$\rm{cts\,s^{-1}}$, corresponding to a 0.5--4\,keV flux of $5\times10^{-10}\,\rm{erg\,cm^{-2}\,s^{-1}}$. As shown in the bottom-left panel of Fig.~\ref{fig:lc}, the hardness ratio (HR; (2--4)\,keV/(0.5--2)\,keV) for both flares reaches its maximum (HR$\sim0.5$), $\sim10-20$\,s before the peak in the count rate, and subsequently decreases during the decay phase.  
After the second flare, no further significant flaring activity was detected at the source position in the subsequent \textit{EP}/WXT observations at the $3\sigma$ confidence level. The average unabsorbed 0.5--4.0 keV flux during the post-flare WXT observations was constrained to an upper limit of  $2.6\times10^{-11}\,\rm{erg\,cm^{-2}\,s^{-1}}$.

Follow-up observations with various X-ray instruments were conducted, and the long-term 0.5--10\,keV X-ray light curve is presented in the right panel of Fig.~\ref{fig:lc}.  Upon its initial detection by \textit{EP}/WXT on MJD 60374.5, EP240305a had an average flux of $7\times10^{-11}\,\rm{erg\,cm^{-2}\,s^{-1}}$ over 17.7 ks (see Tab.~\ref{tab:xray}). Within two days, the flux dropped by two orders of magnitude to $8\times10^{-13}\,\rm{erg\,cm^{-2}\,s^{-1}}$, as observed by \textit{Swift}/XRT, and continued to decrease gradually, reaching an upper limit of $4\times10^{-13}\,\rm{erg\,cm^{-2}\,s^{-1}}$ on MJD 60378.6. The final X-ray observation, conducted with \textit{EP}/FXT on MJD 60395.5, 21 days post-discovery, revealed a flux drop by at least four orders of magnitude, reaching an upper limit of $6\times10^{-15}\,\rm{erg\,cm^{-2}\,s^{-1}}$, by assuming the same spectral shape as \textit{Swift}/XRT.

\subsubsection{Spectral properties}
\label{sec:xray-pha}
For the X-ray spectral analysis, we use {\sc xspec} version 12.13.1 \citep{Arnaud1996}, adopting the abundance table from \citet{Wilms2000} and the cross-section table from \citet{Verner1996}. Due to the low count rates, we analyze the energy spectra only from \textit{EP}/WXT and the first \textit{Swift}/XRT observation. The energy range spans 0.5--4 keV for \textit{EP}/WXT, and 0.5--5 keV for \textit{Swift}/XRT, limited by high background levels beyond 5 keV. We adopt C-statistics \citep{Cash1979} for fitting all spectra, specifying uncertainties for the model parameters at a 90\% confidence level.

\begin{figure*}
\centering
\includegraphics[width=\textwidth]{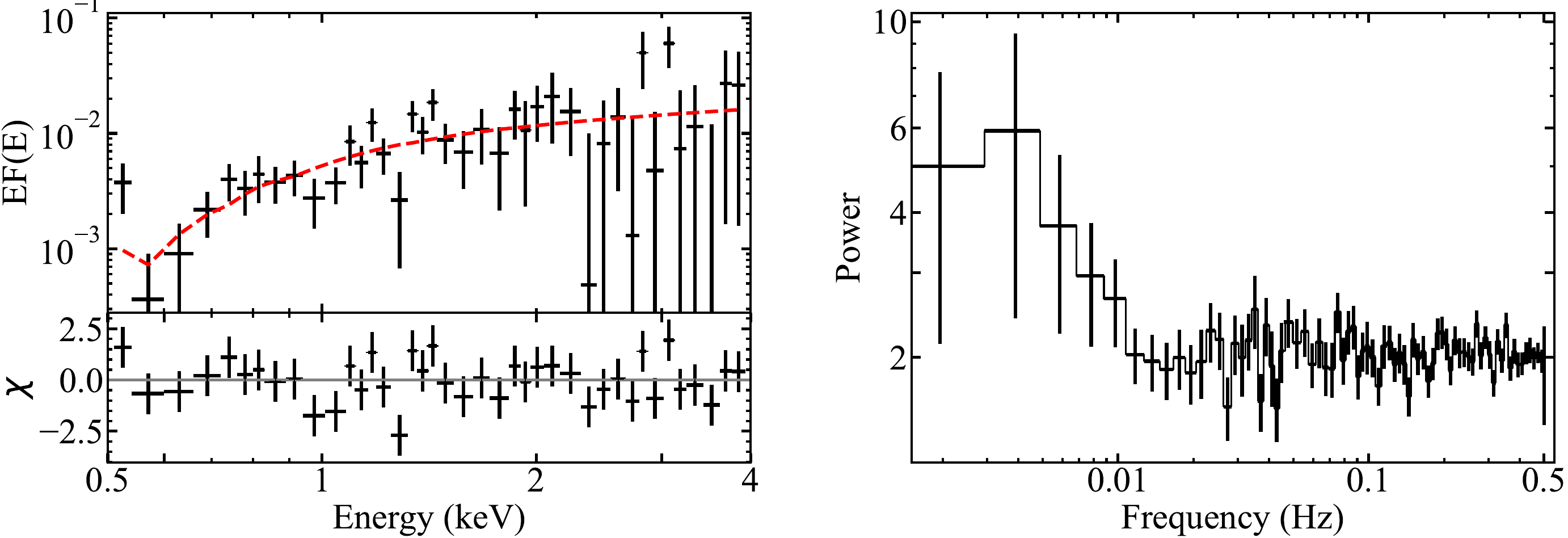} \\
\caption{Left panel: The \textit{EP}/WXT spectrum of EP240305a in the 0.5--4\,keV energy band. The data are best fitted with an absorbed \texttt{powerlaw} model, shown as a red dashed line. The residuals from the best-fit model are displayed in the lower subpanel. Right panel: PDS of EP240305a in the 0.5--4\,keV energy band using \textit{EP}/WXT data. 
\label{fig:wxt_pha_pds}
}
\end{figure*}

We fit the energy spectra from the \textit{EP}/WXT data using an absorbed power-law model \texttt{tbabs*powerlaw}, yielding a C-statistic of 126.16 with 116 degrees of freedom (dof). The fitting results are shown in the left panel of Fig.~\ref{fig:wxt_pha_pds} and Tab.~\ref{tab:xray}. The best-fit photon index, $\Gamma$, is $1.6_{-0.7}^{+0.8}$, indicating a relatively hard spectrum. The absorption column density is measured to be $N_{\rm H}$, is $3_{-2}^{+3}\,\times10^{21}\,{\rm cm^{-2}}$. This value is consistent within the neutral hydrogen column density inferred from the HI4PI H{\sc i} survey \citep{HI4PI2016} ($1.9\,\times10^{21}\,{\rm cm^{-2}}$). It is also in agreement with the independent estimate provided by the 3DN$_{\rm H}$ tool \footnote{\url{http://astro.uni-tuebingen.de/nh3d/nhtool}} ($2.9_{-2.7}^{+2.6}\,\times10^{21}\,{\rm cm^{-2}}$). For the \textit{Swift}/XRT data, $N_{\rm H}$ is not well constrained, and is fixed at the value of $3\,\times10^{21}\,{\rm cm^{-2}}$ from \textit{EP}/WXT. This yields a best-fit C-statistic of 18.3 for 19 dof. The resulting photon index is $\Gamma = 1.2 \pm 0.8$, which supports a hard X-ray spectrum during the outburst.
We further analyzed the spectra of the two individual flares observed by \textit{EP}/WXT. Both are well described by an absorbed power-law model with fixed $N_{\rm H} = 3\,\times10^{21}\,{\rm cm^{-2}}$. The best-fit photon indices are $\Gamma = 1.5_{-0.3}^{+0.4}$ and $\Gamma = 2.0_{-0.5}^{+0.6}$, respectively, 
with no significant spectral variations compared to the entire \textit{EP}/WXT observation.

We also tested an absorbed blackbody model (\texttt{tbabs*bbodyrad}) for the \textit{EP}/WXT spectra. Although the absorbed blackbody model provides statistically acceptable fits, the absorbed power-law model generally yields lower C-statistic values for the whole WXT time-averaged spectrum ($C_{\rm BB}=134.31$ and $C_{\rm PL}=126.16$, with $kT \sim 0.43$ keV), the first flare ($C_{\rm BB}$=28.95 vand $C_{\rm PL}$=26.35, with $kT \sim$0.50 keV), and the second flare ($C_{\rm BB}=23.48$ and $C_{\rm PL}=9.65$, with $kT \sim 0.42$ keV). Thus, while a thermal interpretation cannot be ruled out statistically, the power-law model provides a somewhat better empirical description of the energy spectra.

\subsubsection{Fast time variability}
\label{sec:xray-pds}

We also check the fast timing variability, using only the \textit{EP}/WXT data due to the low statistics of other observations. We use \texttt{powspec} tool in {\sc heasoft} version 6.32 to compute the power density spectrum (PDS) in the energy band 0.5--4\,keV. The time resolution and length of the segment of each Fast Fourier Transformation (FFT) are set as, respectively, 1\,s and 512\,s, corresponding to a lowest frequency of 0.002\,Hz and a Nyquist frequency of 0.5\,Hz for each FFT. The PDS are normalized following the Leahy prescription \citep{Leahy1983}. 

There are no significant quasi-periodical oscillations (QPOs) detected in the PDS; only power-law noise is observed at the low frequencies. We calculated the root mean square (rms) in the 0.5--4\,keV energy band over the frequency range of 0.002--0.5\,Hz, yielding a value of $1.9^{+3.8}_{-1.9}$\%. We note that the WXT light curve is dominated by two short flares; therefore, the averaged PDS mainly reflects the timing properties during these intervals.

\begin{table*}
\centering
\setlength{\tabcolsep}{3pt}
\caption{X-ray observation information and the parameters of the best-fit absorbed power-law model.}
\begin{tabularx}{\textwidth}{XXXXXX}
\hline
Instrument & ObsTime (MJD)  &  Exposure (ks) & $N_{\rm H}$ ($10^{21}\,{\rm cm^{-2}}$)  & $\Gamma$ & Averaged Flux$^{*}$ ($\times10^{-13}\,{\rm erg}\,{\rm cm}^{-2}$\,{\rm s}$^{-1}$) \\ \hline
\textit{EP}/WXT  & 60374.5  & 17.7 & $3_{-2}^{+3}$ & $1.6_{-0.7}^{+0.8}$ & $700_{-200}^{+400}$ \\
\textit{Swift}/XRT  & 60376.7  & 2.7  & 3 (fixed) & $1.2\pm{0.8}$  & $8_{-4}^{+8}$ \\ 
\textit{Swift}/XRT  & 60378.6  &  0.97 & 3 (fixed) & 1.2 (fixed) & <4\\ 
\textit{EP}/FXT  & 60395.5  & 9.5  &  3 (fixed) & 1.2 (fixed) & <0.06 \\
\hline
\end{tabularx}
\footnotesize\raggedright 
Notes. $^*$: Unabsorbed flux in the 0.5--10\,keV band.
\label{tab:xray}
\end{table*}

\subsection{Optical and near-infrared properties}
\label{sec:optical}

To search for a candidate near-infrared counterpart, we performed image differencing on the \textit{GROND} $J$, $H$, and $K_{\rm s}$-band images taken on 2024 March 8th (MJD~60377.102) and 10th (MJD~60379.045), using the \textsc{AutoPhOT} pipeline \citep{Brennan2022d}. For the $J$ and $K_{\rm s}$ bands, reference images from the VISTA Hemisphere Survey\footnote{\url{https://www.eso.org/sci/observing/phase3/data\_releases/vhs\_dr1.html}} obtained in May 2016 were used. For the $H$-band analysis, we used the \textit{GROND} observation from 2024 October 21st (MJD~60379.045) as the reference. For each image, World Coordinate System (WCS) values were verified using \textsc{Astrometry.net} \citep{Lang2010}. An effective point-spread function (ePSF) model was constructed with the \texttt{Photutils} package \citep{Bradley2024}, using bright, isolated sources in each frame. Photometric zero points were calibrated using sequence sources from the ATLAS All-Sky Stellar Reference Catalog \citep[REFCAT2;][]{Tonry2018}, for all three bands.

\begin{figure*}
\centering
\includegraphics[width=0.9\textwidth]{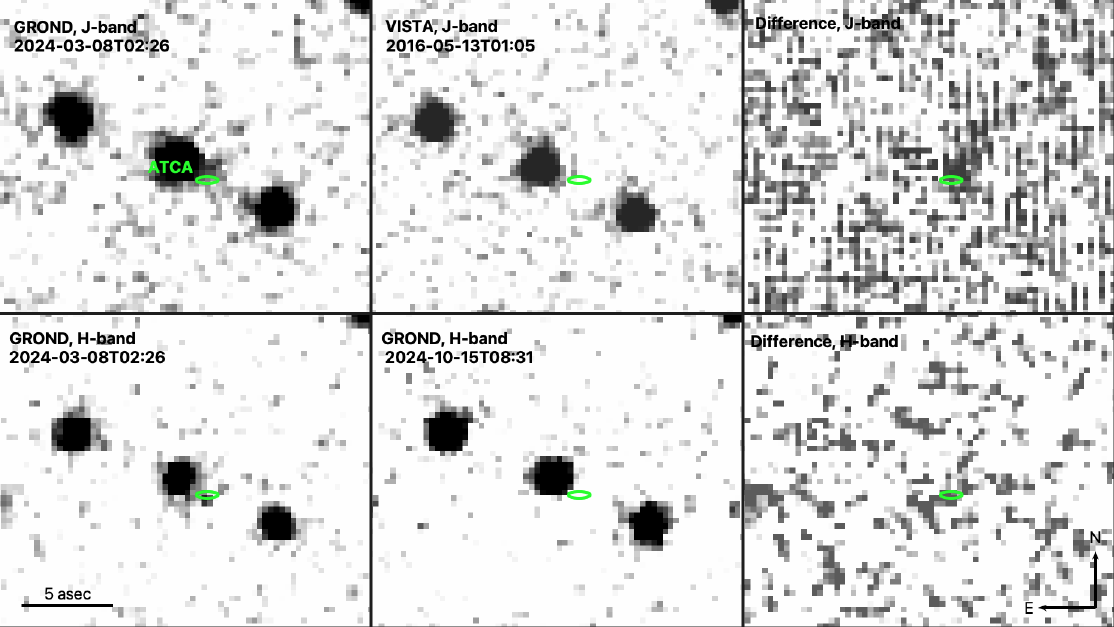} \\
\caption{$J$-band (top row) and $H$-band (bottom row) counterpart analysis. The left and middle columns show the \textit{GROND} images obtained on 2024-03-08 (MJD=60377) and the aligned reference images (VISTA, 2016-05-13 (MJD=57519) for $J$-band; \textit{GROND}, 2024-10-15 (MJD=60598) for H-band), respectively. The right column shows the resulting difference images. The green ellipse indicates the position of the 9\,GHz ATCA radio detection, as observed on MJD=60398 with its uncertainty increased by a factor of 10 in both coordinates to enhance visibility. Signatures of a candidate counterpart are detected in $J$ and $H$, although with a positional offset from each other. See text for more details.
\label{fig:ima_opt}
}
\end{figure*}

The analysis results for the $J$ and $H$ images are shown in Fig.~\ref{fig:ima_opt} and summarised in Table~\ref{tab:grond}. All uncertainties are 1$\sigma$. A faint source (3.4\,$\sigma$ significance) with $J = 19.36 \pm 0.13$ was detected on March 8th and subsequently faded below the detection limit of $J>19.68$ two days later. A source was also detected in the $H$-band on both March 8th (5.3$\sigma$) and 10th (4.1$\sigma$). It showed indications of fading from $H=18.64\pm0.13$ to $18.91\pm0.18$; however, the measurements are also consistent with being constant within the uncertainties. We note that the localisations in the $J$-band (R.A. (J2000) = 08$^{\rm h}$11$^{\rm m}$40.84$^{\rm s}$, Dec. (J2000) = $-54^\circ$39$^\prime$06$\farcs$88, with a positional uncertainty of $0\farcs21$ in both coordinates) and $H$-band (08$^{\rm h}$11$^{\rm m}$40.89$^{\rm s}$, $-54^\circ$39$^\prime$07$\farcs$75, $0\farcs18$) are offset by $0\farcs9$ in Dec. Both positions are broadly consistent with the ATCA radio localisation. 
No source was detected in the $K_{\rm s}$ band down to a limiting magnitude of $K_{\rm s} > 17.2$.

A mid-infrared source, WISE J081140.94$-$543908.1, lies approximately $0\farcs7$ south of the \textit{GROND} position. Given the small positional offset, this source may be associated with the host galaxy of EP240305a. However, the currently available data are insufficient to determine whether it is physically associated with the EP transient.

\begin{table}
\centering
\setlength{\tabcolsep}{3pt}
\caption{GROND Photometry. All magnitudes are given in the Vega system. Uncertainties and upper limits are $1\sigma$.}
\begin{tabularx}{\linewidth}{X X X}
\hline
ObsTime (MJD)  &  Band & Magnitude \\ \hline
60377.102 & $J$ & $19.36\pm0.13$ \\
          & $H$ & $18.64\pm0.13$\\
          & $K_{\rm s}$ & $>17.1$\\
60379.045 & $J$ & $>19.68$ \\
          & $H$ & $18.91\pm0.18$ \\
          & $K_{\rm s}$& $>17.2$\\
\hline
\end{tabularx}
\label{tab:grond}
\end{table}

\begin{figure}
\centering
\includegraphics[width=0.5\textwidth]{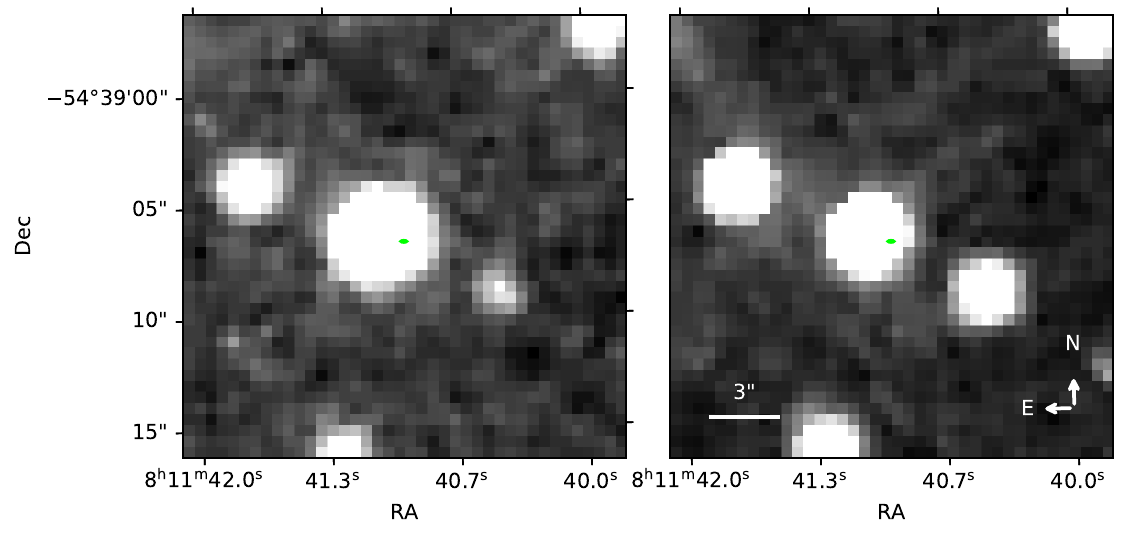} \\
\caption{SVOM/VT observations in the $B$ (left) and $R$ (right) bands. The green dot marks the ATCA position of EP240305a.}   
\label{fig:svom}
\end{figure}

Since EP240305a had already faded by the time of the SVOM/VT observations, we treated the VT image as containing no significant transient emission and first performed standard PSF photometry assuming an isolated point source. To further constrain any possible contribution from a host galaxy, we adopted the source coordinates derived from the ATCA observations and carried out PSF fitting at this fixed position on the residual images. No significant source is detected in the SVOM/VT data, and we derive a 1$\sigma$ upper limit of $R\sim$23.3 mag (see Fig.~\ref{fig:svom}).

\subsection{Radio properties}
\label{sec:radio}

Throughout the observation period, ATCA detected the radio emission from EP240305a up until MJD 60441, noting that the radio source was not spatially resolved. Using the MJD 60398 radio position, which has the smallest positional uncertainty, the angular separations between the \textit{Swift}/XRT position and the radio centroids are $\sim1{\farcs}6$ (5.5 GHz) and $\sim1{\farcs}8$ (9.0 GHz). Both are significantly smaller than the combined positional uncertainty ($\sim5{\farcs}6$), indicating that the radio and Swift/XRT X-ray detections are positionally consistent. Although positional consistency alone cannot establish a physical association, the temporal proximity of the \textit{Swift}/XRT, \textit{GROND}, and ATCA detections to the \textit{EP}/WXT trigger, together with the rarity of unrelated X-ray and radio flaring transients within the \textit{EP}/WXT localization region over this time window, makes a chance coincidence unlikely. We therefore consider the \textit{Swift}/XRT X-ray, \textit{GROND} near-infrared, and ATCA radio detections to be associated with EP240305a.

\begin{figure*}
\centering
\includegraphics[width=0.98\textwidth]{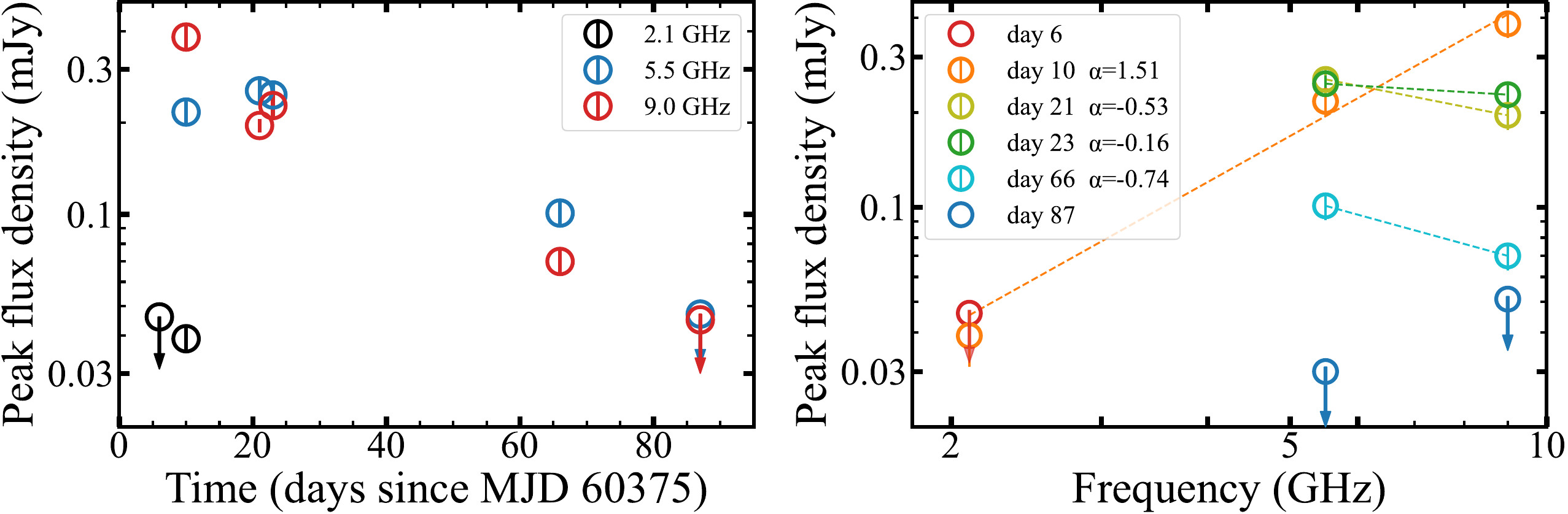} \\
\caption{Left panel: Temporal evolution of flux density observed with ATCA at three distinct frequencies (2.1\,GHz, 5.5\,GHz, and 9.0\,GHz) over the period from March to June, 2024. Right panel: Flux density as a function of frequency for multiple epochs, with spectral indices ($\alpha$) derived from ATCA observations annotated for selected dates.  
\label{fig:radio}
}
\end{figure*}

Fig.~\ref{fig:radio} shows the multi-epoch observations between MJD 60381 and MJD 60462, illustrating the changes in radio flux density across different observing frequencies. The data reveal substantial variability in both flux density and spectral index ($\alpha$), pointing to a complex and evolving jet spectra. The observation from day 6 (MJD 60381) shows an upper limit on the flux density of $<46$\,mJy at the lower frequency of 2\,GHz, representing an early state. On day 10 (MJD 60385), the observation exhibited a strongly inverted radio spectrum with $\alpha = 1.51$\footnote{Owing to the limited frequency coverage, the uncertainty in the spectral index cannot be robustly determined. This limitation similarly affects the indices discussed below.}, suggesting that the emission originated from a very compact, highly self-absorbed synchrotron region at this stage.

Subsequent observations on day 21 and 23 (MJD 60396 and 60398) show a clear shift in spectral behavior, with the spectral index turning negative ($\alpha = -0.53$ and $\alpha = -0.16$, respectively), indicating a transition toward an optically thin regime as self-absorption diminished. This trend continues in the observation on day 66 (MJD 60441), where the spectral index remains negative ($\alpha = -0.74$), consistent with a fully optically thin jet. By day 87 (MJD 60462), the flux density at both 5.5 and 9\,GHz had further declined, and reached the upper limit of 30\,$\mu$Jy and 51\,$\mu$Jy, respectively, suggesting a reduction of jet activity. This spectral evolution—from a self-absorbed to an optically thin state—likely traces the progression of jet ejection and subsequent expansion.

\begin{table*}
\centering
\label{tab:ATCA}

\setlength{\tabcolsep}{3pt}
\renewcommand{\arraystretch}{1.15}
\caption{ATCA observational parameters. }
\begin{tabularx}{\textwidth}{@{}
>{\centering\arraybackslash}p{1.7cm}
>{\centering\arraybackslash}p{4.4cm}
>{\centering\arraybackslash}p{1.7cm}
>{\centering\arraybackslash}p{1.7cm}
>{\centering\arraybackslash}p{1.7cm}
>{\centering\arraybackslash}p{2.8cm}
>{\centering\arraybackslash}p{2.8cm}
@{}}
\hline
MJD & (RA, Dec) & Frequency & Peak flux density & RMS noise level & Beam & Position uncertainty \\
    & (J2000)   & (GHz)     & ($\mu$Jy/beam)    & ($\mu$Jy/beam)  & (arcsec $\times$ arcsec) & (arcsec $\times$ arcsec) \\
\hline
(1) & (2) & (3) & (4) & (5) & (6) & (7) \\
\hline

60381 & ... & 2.1 & $<$46 & 15 & 17.6 $\times$ 6.7 & ... \\
60385 & ... & 2.1 & 39 & 8 & 24.5 $\times$ 2.6 & ... \\

      & 08:11:40.924, $-$54:39:07.628 & 5.5 & 217 & 26 & 12.1 $\times$ 1.6 & 0.7 $\times$ 0.1 \\
      & 08:11:40.893, $-$54:39:07.491 & 9.0 & 383 & 28 & 7.5 $\times$ 1.0 & 0.3 $\times$ 0.04 \\

60396 & 08:11:40.887, $-$54:39:07.375 & 5.5 & 255 & 20 & 6.4 $\times$ 1.3 & 0.2 $\times$ 0.03 \\
      & 08:11:40.854, $-$54:39:07.096 & 9.0 & 196 & 11 & 4.0 $\times$ 0.8 & 0.1 $\times$ 0.02 \\

60398 & 08:11:40.883, $-$54:39:07.627 & 5.5 & 247 & 13 & 3.9 $\times$ 1.3 & 0.1 $\times$ 0.03 \\
      & 08:11:40.877, $-$54:39:07.458 & 9.0 & 228 & 11 & 2.6 $\times$ 0.8 & 0.06 $\times$ 0.02 \\

60441 & 08:11:40.822, $-$54:39:06.759 & 5.5 & 101 & 15 & 5.8 $\times$ 1.3 & 0.4 $\times$ 0.1 \\
      & 08:11:40.883, $-$54:39:07.785 & 9.0 & 70  & 15 & 4.4 $\times$ 0.8 & 0.5 $\times$ 0.08 \\

60462 & ... & 5.5 & $<$30 & 10 & 7.5 $\times$ 5.6 & ... \\
      & ... & 9.0 & $<$51 & 17 & 28 $\times$ 23  & ... \\

\hline
\end{tabularx}
\footnotesize\raggedright 
Notes. Column (1): observation time; Column (2): right ascension and declination coordinates; Column (3): observing frequency; Column (4): peak flux density; Columns (5): root-mean-square (rms) noise of the image; Columns (6): major axis and minor axis of the restoring beam; Columns (7): position uncertainty.
\end{table*}

\section{Discussion and Conclusion} 
\label{sec:disc_conclu}

We analyzed the properties of a newly discovered uncataloged transient source, EP240305a, identified by \textit{EP}/WXT on March 5, 2024 (MJD 60374), across X-ray, optical and radio wavelengths. Subsequent X-ray observations revealed a rapid flux decay, with a two-order-of-magnitude drop within two days. Radio observations, using ATCA over roughly two months, indicate that the radio emission transitioned from being self-absorbed to becoming optically thin. A faint candidate near-infrared counterpart was found at the source position. However, without an optical spectrum, the true nature of EP240305a remains unclear. To better understand this source, we now examine whether its multiwavelength behavior matches that of known transient categories.

\subsection{TDEs and XRB}
\label{sec:disc_XRB}

Given its relatively strong X-ray and radio emission, as well as its long-term radio evolution, we first examine whether EP240305a could originate from a long-timescale event, such as a jetted TDE or an XRB outburst.

Jetted TDEs are a subclass of TDEs exhibiting strong radio emission, produced when the tidal disruption of a star by a supermassive black hole launches a relativistic jet. A well-studied example is Swift J164449.3+573451, which reached an isotropic-equivalent X-ray luminosity of $\sim 10^{47}$\,erg\,s$^{-1}$ and exhibited a $t^{-5/3}$ X-ray decay and suddenly dropped plateau. The radio emission peaks after the X-rays, on timescales of $\sim$100 days. The overall activity lasts for months to years \citep[see review][]{Colle2020}. EP240305a shows markedly different behaviour: its radio emission peaks at $\lesssim$7 days, and its X-ray flux declines by $\gtrsim$2 orders of magnitude within $\sim$2 days and falls below the \textit{Swift}/XRT detection limit within four days. These rapid evolution timescales are incompatible with those of jetted TDEs, effectively ruling out this scenario for EP240305a.

The unusually rapid decay of EP240305a is significantly faster than that observed in typical bright XRB outbursts, which generally reach peak luminosities of $L^{\rm peak}_{\rm X} \sim 10^{37-39}$\,erg\,s$^{-1}$, and persist for several months. This motivates a comparison with a subclass of XRB outbursts characterised by much shorter durations—ranging from several tens of days to a few days \citep{Wijnands2006, Heinke2015, Sguera2024}, referred to as very faint X-ray transients (VFXTs). VFXTs are defined by their low peak X-ray luminosities $L^{\rm peak}_{\rm X} \sim 10^{34-36}$\,erg\,s$^{-1}$ \citep{Degenaar2009,Wijnands2015} and typically exhibit absorbed power-law spectra in the soft X-ray band \citep[e.g.,][]{Shaw2017,Sguera2024}, broadly consistent with the spectral properties of EP240305a. Assuming a VFXT-like luminosity, the inferred distance of EP240305a would be 1--10\,kpc. Given its Galactic latitude $|b| = 11.185^{\circ}$, a Galactic disk origin would imply a relatively smaller distance of $\lesssim$ 5\,kpc. Several observational properties, however, argue against a VFXT interpretation. The source is located at a Galactic longitude of $l = 269.264^{\circ}$, well away from the Galactic centre, and shows a relatively low absorption column $N_{\rm H} \sim 3 \times 10^{21}\,{\rm cm^{-2}}$, both of which are atypical for VFXTs, which are usually concentrated toward the Galactic centre and subject to higher absorption \citep{Muno2005}. Most decisively, VFXTs generally show no detectable radio emission or only very weak radio counterparts \citep{Shaw2017,vdEijnden2021,Sguera2024}. In contrast, EP240305a displays significant radio emission at a level of $\gtrsim$0.3\,mJy, which is difficult to reconcile with a VFXT origin. We therefore suggested that EP240305a is unlikely to belong to the VFXT class.

We also considered the possibility that EP240305a is related to burst-only NS-LMXBs \citep[e.g.,][]{Cocchi2001}. Some fast X-ray transients associated with NS systems have previously been mistaken for GRBs, such as GRB060602B \citep{Wijnands2009}. Although the X-ray spectrum of EP240305a can be fitted with a blackbody model, the temperature ($kT \sim 0.43$ keV) is significantly lower than typical Type-I X-ray bursts. In addition, the long-lasting radio emission and expanding ejecta are not expected from thermonuclear bursts on accreting neutron stars. We therefore consider this scenario unlikely.

The relatively strong and evolving radio emission may instead suggest an association with radio-emitting LMXBs, some of which can exhibit faint X-ray transients accompanied by jet activity, such as CXOGC J174540.0--290031, which has been suggested to be an edge-on BH-LMXB \citep{Muno2005, Bower2005}. However, such systems typically show accretion-driven outbursts lasting for months, together with persistent X-ray activity and orbital modulation signatures. Although the radio evolution suggests ejecta activity, EP240305a decayed by more than four orders of magnitude within only 20 days, making an edge-on BH-LMXB interpretation less likely.

\subsection{Stellar flare and magnetar flare}
\label{dis_short}

Given the short duration of its X-ray flaring activity and relatively low Galactic latitude ($|b| = 11.185^{\circ}$), we consider the possibility that EP240305a is associated with a short-timescale Galactic transient event. One possibility is a stellar flare. Multi-wavelength emission spanning X-ray, optical, and radio bands is commonly observed in stellar flares, which occur frequently, with an estimated all-sky rate of $\sim2.8\times10^6$ events per year \citep{Pye2015}. Among flare stars, M dwarfs are representative, and their flare occurrence rates as functions of spectral type, magnetic activity, and Galactic height have been extensively studied \citep{Kowalski2009}. Assuming a typical M-dwarf absolute magnitude of $M_J \sim 8$--11, the $J$-band limit ($J \sim 20$) implies a distance of approximately 0.6--2.5 kpc, consistent with a Galactic origin. The near-infrared color $J-H = 0.72$ is also characteristic of late-type M dwarfs \footnote{\url{https://www.pas.rochester.edu/~emamajek/memo_colors.html}}.

In addition, M-dwarf flares are known to approximately follow the G\"udel–Benz relation, with $\log(L_{\nu,R}/L_X [\mathrm{Hz}^{-1}]) \sim -15.5$ \citep{Benz1994, Williams2014}). Using quasi-simultaneous ATCA and \textit{EP}/FXT observations obtained on $\sim$MJD 60396, we estimate $\log(L_{\nu,R}/L_X) > -12$ for EP240305a, similar with the population of ultracool dwarfs that deviate from the standard Güdel–Benz relation \citep[see Figure 12 of][]{Williams2014}. However, a stellar-flare interpretation would still require rather extreme conditions. 
If EP240305a were associated with a stellar flare, the observed X-ray fluence ($\sim {\rm (3-5)}\times10^{-7}$\,erg\,cm$^{-2}$) would imply a distance of only $\lesssim 70-100$\,pc, assuming the maximum X-ray energy of known M-dwarf superflares is $\sim 3 \times 10^{35}$\,erg \citep{Joseph2024}. However, no persistent near-infrared counterpart is detected in the $VISTA$ archival data down to $J \sim 20$, corresponding to $M_J \sim 15$ at such distances. This is difficult to reconcile with a typical M-dwarf flare scenario. 
Furthermore, radio emission from stellar flares typically lasts from minutes to hours \citep[e.g.,][]{Tristan2025, Osten2026}, whereas the radio emission associated with EP240305a persists for at least several weeks. 
Taken together, these arguments indicate that a stellar flare origin for EP240305a is unlikely.

Beyond stellar flares, we also considered other short-duration Galactic transients, such as magnetar giant flares. However, the observed behavior is inconsistent with the typical properties of magnetar giant flares, which are characterized by extremely rapid rise times (less than 1\,s), short main peaks (typically $<1$\,s), and extended power-law tails lasting several hundred seconds \citep[e.g.,][]{Hurley1999, Palmer2005, Mereghetti2024}.

\subsection{Gamma-Ray Burst}

More broadly, fast X-ray transients are associated with two main classes of progenitors: stellar flares and GRBs, both of which are expected to be nearly isotropically distributed on the sky \citep{Arefiev2003}. Given the arguments above disfavouring a stellar flare origin, we focus instead on the GRB scenario, motivated by the observed temporal variability in both the X-ray and radio emission.
In the absence of an optical spectroscopic redshift, we first assess whether the luminosities inferred for EP240305a at representative redshifts are physically consistent with a GRB-like extragalactic origin.
Assuming a near- to intermediate-extragalactic distance ($z \sim 0.1-0.5$, $D \sim 0.5-3$\,Gpc), the implied X-ray luminosity and radio luminosity density of EP240305a are $L_X \sim 5.1 \times 10^{44} - 1.9 \times 10^{46}$\,erg\,s$^{-1}$ and $L_{\nu} \sim 1.2 \times 10^{29} - 4.1 \times 10^{30}$\,erg\,s$^{-1}$\,Hz$^{-1}$ (9.0\,GHz), respectively. 
In contrast, a more distant origin ($z > 1$, $D > 6.8$\,Gpc) would imply relatively high X-ray luminosity ($L_X > 9.5\times10^{46}$\,erg\,s$^{-1}$), radio luminosity density exceeding $\sim 2.1 \times10^{31}$\,erg\,s$^{-1}$\,Hz$^{-1}$ (9.0\,GHz).
For comparison, GRB typically exhibit X-ray luminosities of $\sim 10^{46}-10^{48}$\,erg\,s$^{-1}$ at $\sim$1 hr after the burst \citep{Nousek2006}, with median peak radio luminosities of $\sim 10^{31}$\,erg\,s$^{-1}$\,Hz$^{-1}$ at 8.5\,GHz for long GRBs \citep{Chandra2012}. 
Overall, the radio and X-ray properties of EP240305a are broadly consistent with a GRB-like origin, motivating a detailed comparison with GRB features below.

\subsubsection{Comparison with GRB properties}

We first examine the gamma-ray emission associated with EP240305a. Using the \textit{Fermi}/GBM targeted search, \citet{Yadav2025} reported an upper limit of $5.8\times10^{-8}$\,erg\,cm$^{-2}$\,s$^{-1}$ (10--1000\,keV) on the gamma-ray flux and subsequently classified EP240305a as a gamma-ray-dark GRB-like transient. \citet{Zhang2025} reported, using the Multi-Instrument Search method on \textit{Fermi}/GBM and \textit{GECAM}-C data, an upper limit for this source ranging from $1.2 \times 10^{-8}$ to $3.6 \times 10^{-7}\,{\rm erg}\,{\rm cm}^{-2}$\,{\rm s}$^{-1}$ in the 10--1000\,keV band, based on different timescale and spectral profile adopted, without detecting any gamma-ray counterpart. However, it is worth noting that some transients discovered by \textit{EP} have been strongly suggested to be long GRBs, despite showing similarly weak or undetectable gamma-ray emission (see Section ~\ref{dis:compare_GRB} for more details).

The source shows no evidence of QPOs in its PDS (see the right panel of Fig.~\ref{fig:wxt_pha_pds}), aligning with known GRB behavior \citep{Beloborodov1998, Paciesas2012, Guidorzi2012}. Furthermore, the PDS of EP240305a, derived from \textit{EP}/WXT data, follows a power-law distribution, $P(f) \propto f^{-0.8 \pm 0.3}$, at frequencies below 0.01\,Hz. This slope generally matches the typical value of about $-1.0$ reported for the average PDS of 97 GRBs in the same low-frequency range \citep{Guidorzi2012}.

For EP240305a, the \textit{EP}/WXT observation reveals two emission peaks with durations of approximately 120\,s and 250\,s, respectively, separated by a quiescent interval of about 200\,s. 
If these two X-ray flares are associated with activity from a GRB-like transient, their double-peaked structure and intervening quiescent interval resemble the multi-episode emission seen in some long GRBs. Notably, the second flare shows a fast-rise, slower-decay profile similar to typical GRB prompt pulses.
This two-peak structure resembles the 'double burst' pattern observed in GRB 110709B and GRB 121217A—the two most prominent examples—where \textit{Swift}/BAT was triggered twice, with an interval of approximately 10\,mins between the peaks \citep{Zhang2012, Hu2014}. A systematic analysis by \citet{Lan2018} checked the duration of the quiescence phases separating the two peaks in 101 long GRBs observed by \textit{Fermi} (8--1000 keV), and found that quiescent intervals span from $\sim5$ to $\sim220$ seconds. 
Approximately $5$\%--9\% of long GRBs exhibit two-episode emission \citep{Bernardini2013, Hu2014}, indicating that such double peak structures may be a common characteristic in long GRBs.

In the optical, the late-time afterglow (occurring several hours after the trigger) of typical GRBs follows a power-law decay with an index of $\sim -1$ \citep[e.g.,][]{Wang2015, Zhang2018}. For EP240305, the optical counterpart is faint and shows a decaying trend with an approximate slope steeper than $-$0.14 in the $J$ and $H$ bands. However, this estimate is highly uncertain as it is based on only two data points (see Sec.~\ref{sec:optical}). In the radio band, the afterglow of GRBs typically exhibits a peak around 4--6 days, followed by a power-law decay with a slope of $\sim -1$ at 8\,GHz \citep{Chandra2012}. The radio observations of EP240305a began approximately one week after the burst. 
Although we cannot detect the exact date of the flux peak, a fit to the 9\,GHz radio data yields a decay slope of $\sim -0.93\pm{0.07}$, consistent with the expected decline of a radio afterglow \citep{Chandra2012}.

\subsubsection{Comparison with other Gamma-ray-dark GRBs from \textit{EP}}
\label{dis:compare_GRB}

\begin{figure}
\centering
\includegraphics[width=0.5\textwidth]{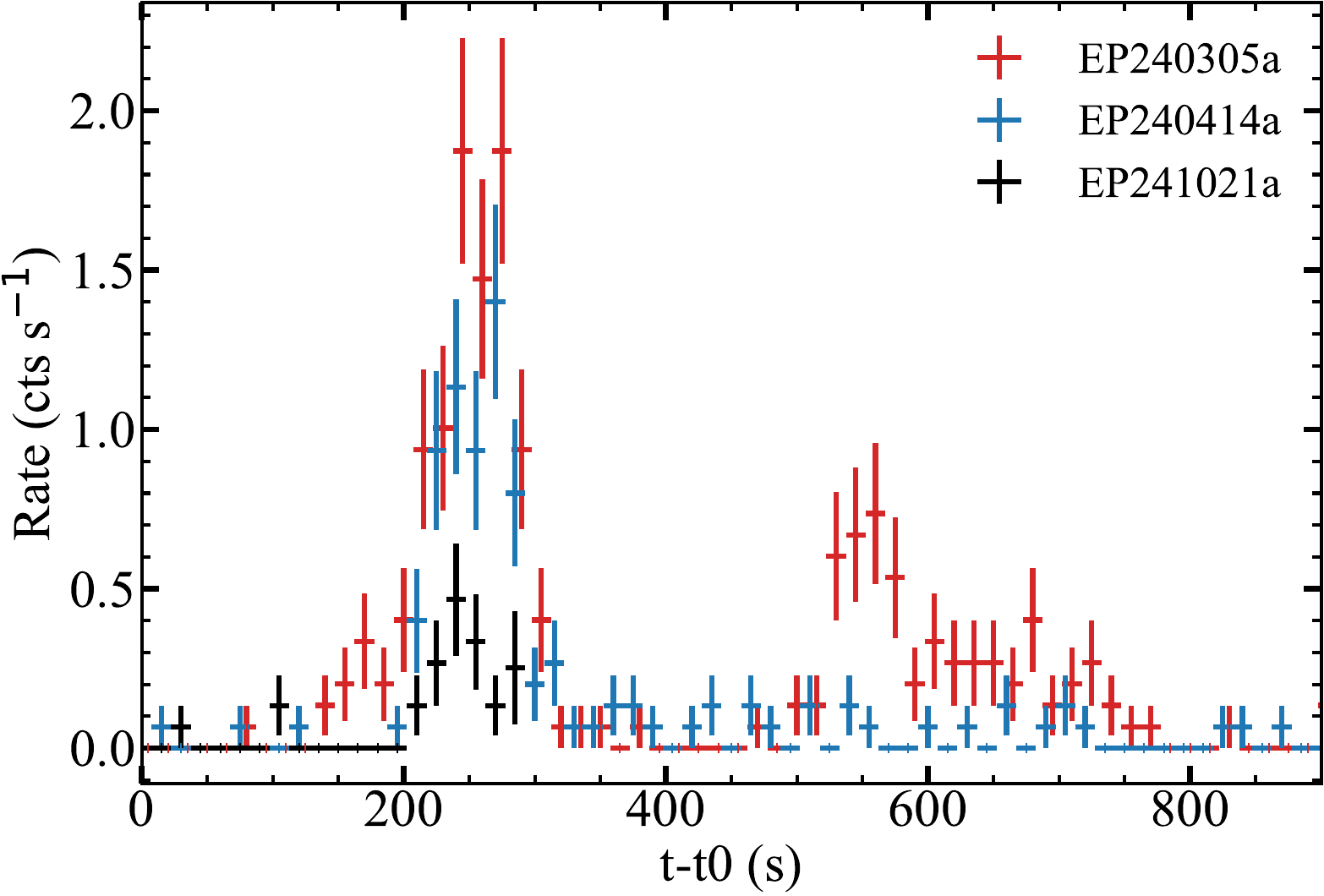} \\
\caption{Comparison of the \textit{EP}/WXT X-ray flares of EP240305a, EP240414a, and EP241021a in the 0.5--4\,keV band. The time axis for each flare has been normalized to a zero point.}   
\label{fig:eps_lcs}
\end{figure}

Notable examples include EP240414a and EP241021a, both initially detected by \textit{EP}, for which redshift measurements and multi-wavelength properties provide strong support for their classification as long GRBs \citep{Bright2025, vanDalen2025, Busmann2025, Sun2025}. 
Another \textit{EP} transient, EP241206a, has also recently been suggested as a possible GRB-like transient occurring within an active galactic nucleus (Wang et al. 2026, in preparation).
To provide a clearer perspective, we compare the properties of these two sources with those of EP240305a, as summarized in Table~\ref{tab:comp} and illustrated in Fig.~\ref{fig:eps_lcs}.

The lack or weakness of gamma-ray emission in such cases may be attributed to several factors \citep{Bright2025, Busmann2025, Yadav2025, Zheng2025}. One possibility is the intrinsic nature of the source, such as low gamma-ray luminosity or low radiative efficiency, which may result from a choked jet or a baryon-rich jet that suppresses efficient gamma-ray emission \citep[dirty fireball;][]{Dermer1999}. Another plausible explanation is a geometric effect, in which the off-axis orientation of the jet leads to a reduced observed gamma-ray flux. 
These events offer valuable opportunities to investigate atypical GRBs with minimal or no gamma-ray output. 
In the case of EP240305a, the current data do not allow us to firmly establish a GRB origin, and we therefore conservatively classify it as a gamma-ray-dark GRB-like transient or or more broadly an extragalactic fast X-ray transient. Nevertheless, its bright X-ray and radio emission likely indicate relativistic jet activity, possibly associated with an off-axis or choked jet \citep[e.g.,][]{Bromberg2012, Bright2025, Ibrahimzade2025}.

Given the sensitivity limits of current X-ray survey satellites, such bursts tend to be overshadowed by more prominent ones. Satellites like \textit{EP}, with high sensitivity and rapid response times, combined with timely multi-wavelength observations, are crucial for probing the properties of these bursts.

\begin{table*}
\centering
\setlength{\tabcolsep}{4pt}
\renewcommand{\arraystretch}{1.25}
\caption{X-ray observation information and the parameters of the best-fit absorbed power-law model. The columns, from left to right, list the source name, X-ray flare duration, X-ray spectral photon index, X-ray flare flux, peak optical brightness reported in the AB system, peak radio flux density, gamma-ray flux upper limit, and references.}
\small
\begin{tabular*}{\textwidth}{@{\extracolsep{\fill}}l c c c c c c c c l}
\hline
Source &
$\Delta t$ &
$\Gamma$ & 
$F_{\rm {x,peak}}$$^a$ &
$M_{\rm o,peak}$ &
$F_{\nu\,\rm {r,peak}}$ &
$F_{\gamma}$ upper limit &
Reference \\
& (s) & & ($\times10^{-9}\,\mathrm{erg\,cm^{-2}\,s^{-1}}$)
& (absolute magnitude) & ($\mu$Jy)
& ($\times10^{-7}\,\mathrm{erg\,cm^{-2}\,s^{-1}}$) & \\
\hline
EP240414a & $\sim150$ & $1.7\pm0.3$ & 3 & $-$21 ($R$) & $434\pm23$ & 2.85--93.7 & 1,2,3,4 \\ 
EP241021a & $\sim100$ & $1.80^{+0.57}_{-0.54}$ & 1 & $-$22 ($R$)  &  $987\pm54$  & 1.35--40 & 5,6,7,8,4 \\ 
EP240305a & $\sim120/250$ & $1.5^{+0.4}_{-0.3}$/$1.0^{+0.6}_{-0.5}$  & 3/0.5 & $-18\sim-24^b$ ($J$)  & $383\pm28$ & 1.2--36  & this work, 4 \\ 
\hline
\end{tabular*}
\label{tab:comp}
\footnotesize\raggedright 
Notes. $^{a}$ X-ray flare flux measured in the 0.5--4\,keV band.\\
$^{b}$ Quantities derived assuming a redshift of $z=0.1-1$.\\
References: (1) \citealt{Lian2024}, (2) \citealt{Srivastav2025}, (3) \citealt{Bright2025}, (4) \citealt{Zhang2025}, (5) \citealt{Shu2025}, (6) \citealt{Hu2024}, (7) \citealt{Busmann2025}, (8) \citealt{Yadav2025}
\end{table*}

\section*{Acknowledgements}

We thank the referee for useful comments that helped
us improve the paper.
We thank Yue Zhao for helpful assistance with the SVOM data.
This work is based on data obtained with the Einstein Probe, a space mission supported by the Strategic Priority Program on Space Science of the Chinese Academy of Sciences, in collaboration with ESA, MPE, and CNES. 
We acknowledge support from China’s Space Origins Exploration Program
Part of the funding for GROND (both hardware as well as personnel) was generously granted from the Leibniz-Prize to Prof. G. Hasinger (DFG grant HA 1850/28-1).
The VISTA Hemisphere Survey data products served at Astro Data Lab are based on observations collected at the European Organisation for Astronomical Research in the Southern Hemisphere under ESO programme 179.A-2010, and/or data products created thereof.
We thank the CSIRO Space and Astronomy staff for supporting these observations. We acknowledge the Gomeroi people as the traditional owners of the Observatory site. 
The Australia Telescope Compact Array is part of the Australia Telescope National Facility (https://ror.org/05qajvd42) which is funded by the Australian Government for operation as a National Facility managed by CSIRO. 

R.M. acknowledges support from the Royal Society Newton Funds. 
SG acknowledges support of the CNES.
F.C.Z. is supported by a Ramóny Cajal fellowship (grant agreement RYC2021-030888-I), by the Spanish grant ID2023-153099NA-I00, and by the program Unidad de Excelencia María de Maeztu CEX2020-001058-M.
J.Z.L. was supported by the Tianshan Talent Training Program through the grant 2023TSYCCX0101, and the Central Guidance for Local Science and Technology Development Fund under No. ZYYD2025QY27.
P.G.J. is supported by the European Union (ERC, Starstruck, 101095973, PI Jonker). Views and opinions expressed are, however, those of the author(s) only and do not necessarily reflect those of the European Union or the European Research Council Executive Agency. Neither the European Union nor the granting authority can be held responsible for them.
X. W. is supported by the National Natural Science Foundation of China (NSFC grants 12288102 and 12033003), the Ma Huateng Foundation, the New Cornerstone Science Foundation through the XPLORER PRIZE. 
Y.X. acknowledges support by National Science Foundation of China through grants NSFC-12521005 and the Hundred Talents Programme of the Chinese Academy of Sciences. R.S. acknowledges hospitality at the Institute of High Energy Physics (Beijing) during part of this work.

\section*{Data Availability}
The reduced data underlying this article will be shared on reasonable request to the corresponding author.



\bibliographystyle{mnras}
\bibliography{ref} 


\noindent\rule[0.5ex]{8.6cm}{0.4pt}
$^{1}$School of Physics and Astronomy, University of Southampton, Highfield, Southampton, SO17 1BJ, UK\\
$^{2}$ Purple Mountain Observatory, Chinese Academy of Sciences, Nanjing 210023, China \\
$^{3}$Key Laboratory of Particle Astrophysics, Institute of High Energy Physics, Chinese Academy of Sciences, Beijing 100049, China \\
$^{4}$Shanghai Astronomical Observatory, Chinese Academy of Sciences, 80 Nandan Road, Shanghai, 200030, China \\
$^{5}$Max Planck Institute for Extraterrestrial Physics, Giessenbachstr. 1, Garching, 85748, Germany \\
$^{6}$INAF-Osservatorio Astrofisico di Torino, Pino Torinese, I-10025, Italy \\
$^{7}$Sydney Institute for Astronomy, School of Physics A28, The University of Sydney, NSW 2006, Australia \\
$^{8}$National Astronomical Observatories, Chinese Academy of Sciences, 20A Datun Road, Beijing 100101, China \\
$^{9}$Shandong Key Laboratory of Optical Astronomy and Solar-Terrestrial Environment, School of Space Science and Physics, Institute of Space Sciences, Shandong \\
University, Weihai, Shandong, 264209, China \\
$^{10}$South African Astronomical Observatory, PO Box 9, Observatory Road, Observatory 7935, Cape Town, RSA \\
$^{11}$Department of Astronomy, University of Cape Town, Private Bag X3, Rondebosch 7701, RSA \\
$^{12}$Department of Physics, University of the Free State, 205 Nelson Mandela Drive, Bloemfontein, 9300, RSA \\
$^{13}$Astrophysics, Department of Physics, University of Oxford, Keble Road, Oxford OX1 3RH, UK \\
$^{14}$Institute of Space Sciences (ICE, CSIC), Campus UAB, Carrer de Can Magrans s/n, Barcelona, E-08193, Spain \\
$^{15}$Institut d’Estudis Espacials de Catalunya (IEEC), Barcelona, E-08034, Spain \\
$^{16}$IRAP, Universit\'{e} de Toulouse/OMP, CNRS, CNES, 9 avenue du Colonel Roche, BP 44346, F-31028 Toulouse Cedex 4, France \\
$^{17}$School of Physics and Astronomy, Sun Yat-sen University, Zhuhai 519082, China \\
$^{18}$CSST Science Center for the Guangdong-Hong Kong-Macau Greater Bay Area, Zhuhai 519082, China \\
$^{19}$Xinjiang Astronomical Observatory, Chinese Academy of Sciences, Urumqi 830011, China \\
$^{20}$Observatoire Astronomique de Strasbourg, Universit\'e de Strasbourg, CNRS, 11 rue de l’Universit\'e , 67000 Strasbourg, France\\
$^{21}$Yunnan Observatories, Chinese Academy of Sciences, Kunming
650216, China\\
$^{22}$School of Astronomy and Space Science, University of Chinese Academy of Sciences, 19A Yuquan Road, Beijing 100049, China\\
$^{23}$Department of Astrophysics/IMAPP, Radboud University Nijmegen, P.O.~Box 9010, Nijmegen, 6500~GL, The Netherlands\\
$^{24}$Physics Department, Tsinghua University, Beijing, 100084, China \\
$^{25}$ School of Astronomy and Space Sciences, University of Science and Technology of China, No. 96 JinZhai Road, Hefei 230026, China \\



\appendix

\counterwithin{table}{section}
\counterwithin{figure}{section}


\bsp	
\label{lastpage}
\end{document}